# Despertando o Interesse de Mulheres para os Cursos em STEM

**Tainara Silva Novaes, Kathleen Danielly Souza Lins, Adolfo Gustavo S. Seca Neto, Mariangela de Oliveira G. Setti, Maria Claudia F. Pereira Emer**

Departamento de Informática
Universidade Tecnológica Federal do Paraná (UTFPR) – Curitiba – PR – Brasil

{kathleenlins.2020, tainaranovaes}@alunos.utfpr.edu.br, {adolfo, mcemer, Mariangelasetti}@utfpr.edu.br,

***Abstract.*** *This article presents initiatives aimed at promoting female participation in STEM fields, with the goal of encouraging more women to pursue careers in these areas. One of these initiatives is the Emílias – Armação em Bits Project, which organizes workshops in schools. Additionally, a podcast has been created to foster interaction between young people and professionals in the field of computing, while also contributing to the establishment of female role models in the industry. The results of these initiatives have been promising, as 70.6% of the students who participated in the workshops expressed an interest in computing. Furthermore, according to Spotify, the podcast's audience consists of 53% females, 44% males, and 3% unspecified, indicating that it has successfully reached a female demographic.*

***Resumo.*** *Este artigo apresenta iniciativas que têm como objetivo promover a participação das mulheres nas áreas de STEM, buscando encorajar mais mulheres a seguirem carreiras nesses campos. O Projeto Emílias - Armação em Bits desenvolve oficinas nas escolas e também um podcast, promovendo a interação entre jovens e profissionais da área de computação, além de contribuir para a formação de modelos femininos nesse campo. Os resultados demonstraram que 70,6% das estudantes demonstraram interesse pela computação após participarem das oficinas. Em relação aos ouvintes do podcast, dados do Spotify indicaram que 53% do público se identifica como feminino, 44% como masculino, e 3% não especificaram o gênero, o que mostra que o podcast tem alcançado um público feminino.*

## 1. Introdução

O tema igualdade de gênero tem sido discutido em diferentes contextos da sociedade e, devido a sua relevância, a Organização das Nações Unidas (ONU) listou a "igualdade de gênero" como um dos 17 objetivos para mudar o mundo [ONU 2015]. No âmbito das universidades, de acordo com estudos apresentados pela ONU Mulheres Brasil (2019), a representatividade das mulheres em áreas como engenharias e tecnologias não chega a 28% do total de estudantes do ensino superior. Esse fato nos mostra a falta de inclusão, interesse ou alta taxa de desistência das mulheres nas áreas STEM (*Science, Technology, Engineering, and Mathematics)*, que podem decorrer de muitos fatores, sejam eles familiares e/ou escolares, como a desigualdade de gênero; cultura machista e/ou patriarcal; sexismo indicado pela discriminação baseada no sexo; negativismo relacionado a aspectos desfavoráveis da cultura e hábitos sociais [Santana e Silva 2020; Souza e Loguercio 2021].

Estudos realizados apontam fatores importantes que permeiam a baixa representatividade das mulheres em áreas da computação e mostram a necessidade de que ações continuem a ser realizadas para que essa realidade seja alterada e possibilite a equidade de gênero nessa área [Ramos e Araújo 2022; Medeiros et al. 2022; Ribeiro et. al. 2019]. Nesse contexto, o projeto de extensão *Emílias – Armação em Bits* da *Universidade Tecnológica Federal do Paraná (UTFPR)* dissemina conhecimento sobre computação para as mulheres, apresentando possibilidades de atuação em diferentes áreas do mercado.

O objetivo deste projeto, abordado neste artigo, é atrair as estudantes do ensino fundamental e médio de escolas públicas para cursos de computação por meio de atividades como oficinas, eventos, palestras informativas e/ou motivacionais, e podcasts. As oficinas realizam atividades, em sua maioria, desplugadas para introduzir conceitos da computação às estudantes. O *podcast* evidencia a representatividade feminina na área de Computação e em setores correlatos, entrevistando mulheres que atuam na área e contam sobre sua trajetória acadêmica e profissional.

Além disso, o projeto fornece suporte e incentivo às alunas dos cursos de graduação em computação da universidade, a fim de mantê-las nos cursos e reduzir a evasão. Isto porque, apesar de um movimento crescente do gênero feminino, nas universidades o número de mulheres em cursos de STEM permanece reduzido [Arruda et al. 2021]. Esta realidade pode ser observada na *UTFPR*, a partir de informações coletadas na base de dados da universidade, que evidenciam uma baixa representatividade de mulheres em formação (estudantes regulares) ou graduadas (estudantes egressas) na área da computação, conforme pode ser visto na Tabela 1.

**Tabela1: Quantidade de Estudantes Regulares e Egressos**

| Cursos/ período | Quantidade de Estudantes Regulares | | | | Quantidade de Estudantes Egressos | | | |
|---|---|---|---|---|---|---|---|---|
| | 01/09 a 02/15 | | 01/16 a 02/22 | | 01/09 a 02/15 | | 01/16 a 02/22 | |
| | Total | Mulheres (percentual) | Total | Mulheres (percentual) | Total | Mulheres (percentual) | Total | Mulheres (percentual) |
| Sistemas de Informação | 300 | 45 (15%) | 436 | 68 (16%) | 43 | 8 (19%) | 163 | 32 (20%) |
| Engenharia da Computação | 353 | 38 (11%) | 435 | 49 (11%) | 66 | 7 (11%) | 210 | 28 (13%) |
| Engenharias | 2932 | 542 (18%) | 4027 | 800 (20%) | 247 | 57 (23%) | 2284 | 507 (22%) |

Fonte: Autoria própria (2022).

É importante observarmos, segundo os dados da Tabela 1, que a partir de 2016 houve um acréscimo no número de graduandas e egressas. Acreditamos que o projeto contribuiu para esse aumento; inclusive tivemos relatos de estudantes que participaram de nossas oficinas no ensino médio e agora participam de nossas atividades como estudantes de graduação, atuando como voluntárias do projeto.

O artigo está organizado como segue: contexto; ações do projeto, apresentação das atividades realizadas; resultados e discussões, relacionados a execução das oficinas, métricas do *podcast* e demais atividades; por fim, considerações finais e agradecimentos.

## 2. Contexto

Nos últimos anos, tem sido amplamente debatida a questão da presença feminina nas áreas de tecnologia e ciências, em relação à representatividade dessas profissionais no mercado. Embora a busca pela equidade de gênero seja uma meta a ser alcançada em diversas áreas, as que envolvem a tecnologia e as ciências são frequentemente mencionadas como particularmente desiguais [Sampaio et. al. 2020]. Vários fatores contribuem para a falta de representação feminina nessas áreas, incluindo estereótipos de gênero, falta de mentoria, discriminação no ambiente de trabalho e barreiras sistêmicas que impedem o acesso e a ascensão de mulheres [Souto et. al. 2022]

Segundo Ramos e Araújo (2022), que realizaram uma pesquisa com estudantes de cursos de computação, entre os fatores que desmotivam as mulheres a continuarem cursando a graduação estão o preconceito contra o gênero feminino, piadas depreciativas feitas por colegas de curso e também professores, bem como a baixa representatividade feminina no curso e no mercado de trabalho. Já no estudo realizado por Medeiros et al. (2022), que abordou a percepção das alunas de ensino médio da rede pública sobre a área de TI, os autores concluíram que apesar da área da computação ser vista como uma boa escolha profissional, diversas barreiras afastam essas alunas, entre elas: a falta de apoio familiar e estrutural, os estereótipos e estigmas machistas que cercam a área e as dificuldades enfrentadas no ensino superior e no mercado de trabalho.

Apesar dos esforços de muitas organizações e indivíduos para aumentar a participação das mulheres na ciência e tecnologia, ainda há muito a ser feito para melhorar a representatividade feminina nessas áreas. Programas de mentoria e educação têm sido criados com o objetivo de incentivar meninas e mulheres jovens a seguirem carreiras nessas áreas, além disso, diversas empresas estão trabalhando para implementar políticas de inclusão e aumentar a diversidade em suas equipes. De acordo com Ribeiro et. al. (2019), é fundamental promover discussões e reflexões contínuas sobre o tema, estendendo-se também para outras áreas, a fim de buscar soluções e avançar na busca por equidade de gênero. Ainda, os autores notaram que, embora os dados de sua pesquisa mostrem diferenças significativas relacionadas ao gênero (ingressos, egressos e média salarial), há uma crescente conscientização acerca do tema e da necessidade de transformar essa realidade.

Portanto, ainda é necessário um grande esforço para combatermos o desequilíbrio de gênero na STEM. É importante continuar criando oportunidades educacionais e programas de mentoria para meninas e mulheres, além de promover modelos femininos de sucesso nessas áreas. Aumentarmos a representatividade feminina na ciência e tecnologia é essencial não apenas para alcançar a equidade de gênero, mas também para fomentar a inovação e encontrar soluções mais criativas e eficazes para os desafios enfrentados pela sociedade atualmente [Machado et. al. 2020].

### 2.1. Trabalhos relacionados

Na literatura podemos observar artigos que tratam de iniciativas que se alinham as ações realizadas pelo *Emílias - Armação em Bits*, entre eles, Lima et. al. (2022), Hoger et. al. (2022), Crestani et. al. (2019). Os artigos trazem estudos que abordam a importância de realizar atividades para estudantes do ensino fundamental e médio, de modo a despertar o interesse das mulheres pela área da computação. Entre as atividades, segundo Hoger et. al. (2022) os artigos apresentam estratégias para tornar oficinas de STEM mais efetivas e engajadoras para as alunas. Há trabalhos que avaliam o impacto das oficinas de STEM e computação nas habilidades e competências dos alunos em geral, como a resolução de problemas, pensamento crítico e criatividade. Ainda é possível analisarmos a relação entre a participação em oficinas de STEM e o desempenho acadêmico dos alunos em disciplinas relacionadas [Cardoso 2020]. Além disso, podemos encontrar estudos que investigam o uso de *podcast* [Quadros 2019].

O artigo de Lima et. al. (2022) adota uma perspectiva diferente, direcionada para as docentes da rede estadual do Mato Grosso. O estudo busca auxiliar as professoras do ensino fundamental a incluir assuntos de STEM e gênero em suas metodologias de ensino na escola. Já em Hoger et. al. (2022), são explorados, de forma apurada e sensível, os possíveis erros cometidos em oficinas que possam reforçar estereótipos sexistas existentes, mostrando a importância de criar estratégias para evitá-los.

Em se tratando de *podcasts*, Crestani et. al. (2019) afirmam que os *podcasts* auxiliam a disseminar informações de maneira mais descontraída e interativa, podendo ser utilizados como uma ferramenta de divulgação científica, seja levando informações de especialistas do meio acadêmico ao público em geral, seja mostrando o que é feito dentro da universidade aos jovens que tenham interesse pela área vinculada com o *podcast* e que desejem ingressar na universidade.

Em resumo, há uma variedade de trabalhos que mostram a importância e os benefícios das oficinas nas escolas para áreas de STEM e computação, bem como exploram diferentes abordagens para tornar essas atividades mais efetivas e acessíveis para as alunas.

## 3. Ações do Projeto

As oficinas propostas foram executadas nas escolas por meio de atividades como "Prototipando Ideias", "*Emílias* no País do Banco de Dados", "Jogos com as *Emílias*", entre outras. Para a realização das oficinas é necessário que haja uma reunião com a diretoria do colégio para explicação das atividades propostas, definição de horários, salas ou laboratórios, local para atividade ser realizada (na escola ou na *UTFPR*) e quantidade de vagas para participação. Além dessas oficinas, o projeto também realiza outras, como a oficina de *Software* Livre, Arduino e *Scratch*.

"Prototipando Ideias" é uma oficina desplugada, sem o uso do computador, na qual a turma de alunas é dividida em grupos para desenvolverem um protótipo em papel de um aplicativo que aborde temas como: saúde, meio ambiente, educação, segurança, mobilidade urbana ou problemas sociais. No início da atividade, exemplos de aplicativos criados por adolescentes são apresentados para inspirar as alunas. Alguns desses exemplos são obtidos do evento anual *Technovation Girls*[1], que é um exemplo prático dessa atividade, na qual equipes de meninas, de diferentes partes do mundo, participam para aprender a resolver problemas do mundo real por meio de soluções tecnológicas.

---
[1] https://www.technovationbrasil.org/

A oficina "*Emílias* no País do Banco de Dados" é realizada em um laboratório de informática e apresentamos para as estudantes a importância de banco de dados (BD), os conceitos básicos e as diferentes formas de uso de um BD, como por exemplo, em bibliotecas digitais. Em seguida as estudantes são divididas em grupos para entender na prática como um banco de dados é implementado, elas são desafiadas a preencher e ordenar dados em uma planilha compartilhada, para que percebam a importância da comunicação do trabalho em grupo e de manter padrões para organização de dados, bem como, entendam conceitos de compartilhamento e concorrência.

"Jogos com as *Emílias*" é uma atividade em que as(os) estudantes conhecem um pouco a respeito da trajetória e descobertas de cientistas de diferentes áreas do conhecimento por meio dos jogos de memória, dama chinesa e pescaria. Nessa atividade participam as alunas e os alunos, porque é preciso mostrar para todos o impacto, os avanços e contribuições que muitas mulheres fizeram e fazem para a área de STEM. Com a realização dessa atividade pretendemos mostrar figuras femininas, exemplos de mulheres de sucesso, que possam ser identificadas pelas estudantes como referências, e além disso, permitir o início de uma discussão sobre os motivos que levaram essas mulheres a invisibilidade.

No *podcast* as entrevistas têm como finalidade conhecer mais sobre o percurso acadêmico, profissional e pessoal de mulheres atuantes na área da Computação. Nessas entrevistas é fundamental abordar as dificuldades e as motivações das mulheres em suas trajetórias. Dessa foram, pretendemos mostrar exemplos e incentivar outras mulheres a adentrarem na área da tecnologia. Antes do período pandêmico de 2020 e 2021, as entrevistas do *podcast* aconteciam preferencialmente na sala de audiovisual do Campus Curitiba da *UTFPR*. Entretanto, com as adaptações que foram feitas por conta da pandemia, o *podcast* começou a realizar entrevistas de forma virtual. Este caminho, que tinha sido pouco explorado anteriormente, permitiu que o *podcast* fosse muito além das fronteiras de Curitiba e do Brasil. Em 2022, o *podcast* retomou as gravações no estúdio; contudo, entrevistas remotas ainda são realizadas, a depender da localização e disponibilidade da entrevistada.

## 4. Resultados e Discussões

### 4.1 Oficinas nas Escolas

Em 2022, após o período de pandemia, retomamos as oficinas presenciais, a primeira delas intitulada como "Prototipando Ideias", ministradas em colégios públicos da região metropolitana de Curitiba, atingindo um público de aproximadamente 400 estudantes, com idades entre 13 e 17 anos. Em agosto do mesmo ano, foi apresentada a oficina "*Emílias* no País do Banco de Dados" para as mesmas turmas dos colégios. Todas as oficinas foram executadas com o apoio de estudantes bolsistas e voluntárias(os) do projeto.

Durante a realização das oficinas, aplicamos questionários para descobrir o perfil das estudantes, suas percepções sobre a área de tecnologia e também para obter *feedback* sobre as atividades. A aplicação de dois questionários, um no início e outro ao final da atividade, se mostrou uma estratégia eficaz para entendermos e captarmos as necessidades e o cenário da realidade das participantes. Com base nos dados coletados durante as oficinas, foi possível observar um crescente interesse das estudantes pela área de computação após a participação em atividades relacionadas à TI. Mais especificamente, 70,6% se interessaram pela computação após as atividades realizadas. Ainda, 79% das alunas apresentaram interesse em fazer um curso de graduação na área, mas não souberam responder qual curso desejavam realizar.

O uso dos questionários permitiu a avaliação dos resultados alcançados e a comprovação da importância dessas ações. Com isso, os projetos podem ser ajustados e direcionados de maneira mais adequada para atender às necessidades e demandas dessas jovens, tornando as ações ainda mais efetivas. Um dado curioso observado a partir dos questionários aplicados nas oficinas foi que, quando as jovens foram questionadas se conheciam alguém que estudava ou trabalhava na área de computação, mais da metade confirmou que sim. Entretanto, quando foi solicitado que elas descrevessem a área de atuação da pessoa conhecida, elas tiveram dificuldades para explicar o que essas pessoas faziam.

Essa dificuldade pode estar relacionada ao fato de que, muitas vezes, as mulheres são sub-representadas na computação, o que leva a uma falta de referências femininas na área, ou ainda, de que por se tratar de uma área vista, muitas vezes, como masculina, gera a falta de interesse sobre as oportunidades de atuação nessa área. Dessa forma, mesmo que as jovens conheçam alguém que estude ou trabalhe em computação, essas referências podem não ser suficientes para que elas tenham uma ideia clara sobre a área de atuação. Tal escassez de referências pode ser um obstáculo para que mais mulheres se interessem por essa área, e é por isso que iniciativas como o projeto *Emílias - Armação em Bits* são tão importantes. Ao promover o contato direto das jovens com estudantes e profissionais de computação, estamos contribuindo para a formação de referências femininas na área, incentivando e inspirando mais mulheres a seguirem carreiras nestes setores.

Dessa forma, a observação da dificuldade das jovens em descrever a área de computação, mesmo conhecendo alguém que trabalhe ou estude na área, reforça a importância de iniciativas que visem a inclusão e a representatividade feminina nas áreas de STEM, como forma de ampliar o interesse e a participação de mulheres nessas áreas.

Outro ponto importante da realização dessas oficinas é a colaboração entre universidade e colégio, que contribui para troca de conhecimentos e experiências entre estudantes e docentes das duas instituições, enriquecendo o processo de aprendizagem e oferecendo a visão de novas possibilidades para as estudantes.

É notável a complementaridade das iniciativas de oficinas realizadas pelo *Emílias* para estudantes e as oficinas para professores, conforme discutido por Lima et al. (2022) e Hoger et al. (2022). Ao capacitar os professores para incorporarem STEM no ensino desde as séries iniciais, desperta-se o interesse dos alunos em fases cruciais de sua formação. Além disso, as oficinas de STEM direcionadas às jovens estudantes possibilitam que elas compreendam que a ciência e a tecnologia são campos acessíveis e realizáveis por mulheres. Dessa forma, é possível atenuarmos essa disparidade de gênero de forma mais ágil, combinando ações direcionadas a esse grupo e corrigindo o problema em sua raiz.

**4.2 *Podcast***

O *Emílias Podcast - Mulheres na Computação* busca entrevistar, majoritariamente, meninas e mulheres que atuam nas mais diferentes subáreas da computação. Entre as perguntas que fazemos com frequência estão:
- Como você se interessou pela área da Computação?
- Como foi sua formação na área?
- Você enfrentou dificuldades, na escola ou no trabalho, por ser mulher?
- Alguma mulher te inspirou em sua carreira?
- O que você diria para meninas ou mulheres que estejam pensando em seguir carreira na Computação?

Para conhecer melhor o público do *podcast*, aplicamos um formulário digital e obtivemos respostas que descrevem resultados interessantes. A nossa pesquisa buscou compreender a eficácia da divulgação do projeto, bem como a disposição das pessoas em participar como entrevistadoras ou entrevistadas. Além disso, coletamos sugestões e elogios dos respondentes.

A partir dessa amostra, buscamos entender se os objetivos do projeto estavam sendo alcançados. Avaliamos a divulgação do projeto e identificamos que as estratégias de publicidade utilizadas foram relativamente eficazes, pois o *podcast* foi conhecido por meio de diversas formas. Mais de 50% dos respondentes mostraram disposição em participar como entrevistadoras ou entrevistadas. Algumas mulheres utilizaram a opção de relatos para compartilhar suas experiências de participação como entrevistadas no *podcast*. Recebemos sugestões e elogios, incluindo elogios ao trabalho da equipe, à diversidade de temas abordados, à importância de trazer a visibilidade das mulheres na computação para a universidade e para as redes sociais. Outro ponto destacado pelos respondentes foi a necessidade de continuidade e crescimento do *podcast*.

Além da pesquisa via formulário, realizamos uma análise dos dados fornecidos pelas plataformas de *streaming* Spotify *for Podcasters* e YouTube para avaliar o alcance do projeto. O canal do YouTube contou com 381 perfis inscritos até março de 2023. Porém, 78,3% do tempo de exibição de vídeos vem de perfis não inscritos no canal. Em relação às visualizações, 55,7% vêm de espectadores do gênero feminino e 44,3% do masculino. Em toda a história do canal no YouTube tivemos 13.815 visualizações de nossos vídeos, que são majoritariamente do *podcast*, mas também incluem eventos, palestras e outros tipos de vídeo. Na Spotify *for Podcasters*, plataforma de streaming de áudio que utilizamos para publicar nossos episódios, contabilizamos 7.955 reproduções de episódios, com uma média de 21 reproduções por episódio. No Spotify, plataforma usada por 31,7% de nosso público ouvinte, 52,3% do público se identifica como feminino, 44,6% como masculino, 2,8% é considerado não especificado e 0,3% não-binário.

A análise dos dados fornecidos pelas plataformas de streaming revela que o projeto tem crescido em número de inscritos e visualizações, especialmente entre o público feminino. Isso sugere que o *podcast* tem alcançado seu público-alvo e contribuído para apresentar a computação e áreas correlatas para meninas e mulheres.

**4.3 Acolhimento**

O projeto *Emílias - Armação em Bits* também promove e organiza eventos anuais, sendo o primeiro deles o Dia Internacional da Mulher, no qual palestras e oficinas são oferecidas para a comunidade interna e externa. As palestras são conduzidas por mulheres, com o objetivo de reforçar a importância da comemoração da data. Além disso, é fundamental dar espaço e evidenciar a importância de amplificar a voz das mulheres.

Recepção aos calouros são promovidas na universidade com o objetivo de receber os novos estudantes no ambiente universitário. O *Emílias* participa dessa recepção se aproximando principalmente das alunas dos cursos de TI, que frequentemente se deparam com ambientes predominantemente masculinos na graduação, e também para apresentar o projeto aos ingressantes. Além disso, outros eventos são realizados com as estudantes durante cada semestre, como sessões de filmes seguidas de roda de conversa, *coffee breaks*, oficinas e outras atividades.

O Ada Lovelace Day é outro evento anual celebrado e organizado pelo *Emílias - Armação em Bits* na universidade. Esse evento é realizado mundialmente na segunda terça-feira de outubro de cada ano em homenagem a Ada Lovelace, considerada como a primeira pessoa programadora do mundo. O objetivo do evento é aumentar a conscientização sobre a importância da diversidade de gênero na computação, bem como inspirar e motivar mulheres e meninas a seguir carreiras em STEM. Nesse evento são realizadas palestras técnicas, rodas de conversa e oficinas variadas.

## 5. Considerações Finais

O projeto de extensão visa promover a equidade de gênero e a inclusão das mulheres na área de STEM, por meio das oficinas nas escolas e do *podcast*. Os dados obtidos das respostas aos formulários, da análise de plataformas de *streaming* e dos relatos de participantes em ações realizadas indicam que o projeto está alcançando seus objetivos e contribuindo para a formação de uma sociedade mais consciente e igualitária.

As oficinas nas escolas têm realizado atividades desplugadas ou não, que visam incentivar a participação feminina na área da computação e correlatas, despertando o interesse das estudantes em aprender mais sobre essa área. O *podcast* tem sido disseminado de forma ampla, atingindo pessoas por meio de diversas plataformas, de modo que os números de inscrições e visualizações no YouTube, Spotify e Anchor indicam um crescimento contínuo do projeto. Além disso, mais da metade dos respondentes do formulário apresentaram disposição para atuar como entrevistadora ou entrevistada, o que sugere que o projeto está contribuindo para a apresentação e desmistificação da computação para as meninas e mulheres.

Assim sendo, podemos dizer que o projeto está contribuindo para a apresentação da computação para estudantes do ensino fundamental e médio, mostrando oportunidades e inspirando meninas e mulheres, a partir de exemplos de mulheres de sucesso que atuam na área da computação na academia e no mercado, incentivando a participação e o aumento da representatividade feminina em STEM, e consequentemente, promovendo a equidade de gênero e o desenvolvimento de uma sociedade mais inclusiva. Entretanto, apesar de que o Emílias tenha estado presente desde 2013, não podemos afirmar de maneira conclusiva que houve um aumento estatisticamente significativo no número de mulheres matriculadas em nossos cursos.

## 6. Agradecimentos

Aos colégios participantes, agradecemos pela oportunidade concedida às jovens garotas de adquirir conhecimentos sobre a carreira de TI. Expressamos nossa gratidão às profissionais entrevistadas no podcast, que generosamente compartilharam suas experiências pessoais e profissionais na busca pela igualdade de gênero. Também estendemos nosso reconhecimento às empresas parceiras que apoiam iniciativas e programas direcionados a mulheres na área de TI, bem como à Fundação Araucária, que concede bolsas aos estudantes e contribui com o projeto. À UTFPR, nosso sincero agradecimento por ceder espaço para a realização das atividades. Por fim, agradecemos à comunidade, que interage, apoia e demonstra cada vez mais interesse nesse importante tema.